\documentclass[pra,aps,twocolumn,showpacs,superscriptaddress]{revtex4}

\usepackage{amsmath}
\usepackage{graphicx}
 
\begin{document}
 
\title{Statistical properties of the continuum Salerno model}

\author{M. Marklund}
\altaffiliation[Also at: ]{Centre for Fundamental Physics, Rutherford Appleton Laboratory,
  Chilton, Didcot, Oxon OX11 OQX, U.K.}
\affiliation{Centre for Nonlinear Physics, Department of Physics, 
  Ume{\aa} University, SE--901 87 Ume{\aa}, Sweden}

\author{P. K. Shukla}
\altaffiliation[Also at: ]{Centre for Nonlinear Physics, Department of Physics, 
  Ume{\aa} University, SE--901 87 Ume{\aa}, Sweden}
\altaffiliation{Centre for Fundamental Physics, Rutherford Appleton Laboratory,
  Chilton, Didcot, Oxon OX11 OQX, U.K.}
\altaffiliation{Department of Physics, University of Strathclyde, Glasgow, Scotland, 
  G4 ONG, UK}
\affiliation{Institut f\"ur Theoretische Physik IV and Centre for Plasma Science 
  and Astrophysics, Fakult\"at f\"ur Physik und Astronomie, Ruhr-Universit\"at Bochum, 
  D--44780 Bochum, Germany}

\author{R. Bingham}
\affiliation{Space Science \& Technology Department, Rutherford Appleton Laboratory,
  Chilton, Didcot, Oxon OX11 OQX, U.K.}

\author{J. T. Mendon\c{c}a}
\altaffiliation[Also at: ]{Centre for Fundamental Physics, Rutherford Appleton Laboratory,
  Chilton, Didcot, Oxon OX11 OQX, U.K.} 
\affiliation{Centro de F\'{i}sica dos Plasmas, Instituto Superior T\'{e}cnico, 1049-001 Lisboa, 
  Portugal}
  
\date{Received 10 April 2006}

\begin{abstract}
  The statistical properties of the Salerno model is investigated. 
  In particular, a comparison between the coherent and partially
  coherent wave modes is made for the case of a random phased wave
  packet. It is found that the random phased induced spectral
  broadening gives rise to damping of instabilities, but also
  a broadening of the instability region in quasi-particle 
  momentum space. The results can be of significance 
  for condensation of magnetic moment bosons in deep
  optical lattices.  
\end{abstract}
\pacs{05.45.Yv, 63.20.Ry, 63.20.Pw}

\maketitle

\newpage

Nonlinear wave propagation models are of importance in a wide variety 
of settings, such as in nonlinear optics \cite{kivshar-agrawal,shukla-rasmussen},
in water wave propagation \cite{onorato-etal}, and in plasma systems \cite{hasegawa}.
Moreover, also discrete systems show important nonlinear properties that
give rise to solitary structures in e.g.\ lattices (see, e.g. \cite{hennig-tsironis} 
and references therein for a review). The discrete Salerno model \cite{salerno} combines 
onsite and intersite nonlinear terms in a lattice in order to encompass the different 
characteristics of nonlinear integrable and non-integrable lattice models. It has also 
been discovered as an asymptotic form of the Gross--Pitaevskii equation, with applications 
to Bose--Einstein condensation of atoms with magnetic moments in deep optical lattice 
traps \cite{li-etal,gomez-gardenes-etal}. Recently, the continuum limit of the Salerno 
model with focusing onsite and defocusing intersite nonlinearities 
was investigated \cite{gomez-gardenes-etal} and soliton solutions were found. 

In this Brief Report, we will investigate the modulational instability properties of the
general continuum limit of the Salerno model. Furthermore, the effects of
random phased wave packets on the stability properties will be studied
using the Wigner formalism. Results regarding the effect of random pulse phases 
will be given.

The Salerno model was derived as a quantum modified discrete nonlinear Schr\"odinger equation, giving the time evolution of the field amplitude on the lattice. The relevant equation takes the form \cite{salerno}
\begin{equation}
  i\partial_t\Phi_n + (\Phi_{n+1} + \Phi_{n-1})(1 + \mu|\Phi_n|^2) + 2\nu|\Phi_n|^2\Phi_n = 0 ,
\end{equation}
where $\Phi_n = \Phi(t,x_n)$ is the field amplitude on the $n$th lattice site located at $x_n$ and 
$\nu$ and $\mu$ are the real parameters determining the strength of the onsite and 
intersite nonlinearities, respectively. The continuum approximation is introduced by the ansatz $\Phi(t,x) = \Psi(t,x)\exp(2it)$ and using the expansion $\Psi_{n\pm1} \approx \Psi \pm \partial_x\Psi +\partial^2_x\Psi/2$ 
the continuum limit of the Salerno model is obtained from Eq.\ (1) according to \cite{gomez-gardenes-etal}
%1
\begin{equation}\label{eq:salerno}
  i\partial_t\Psi + 2(\nu + \mu)|\Psi|^2\Psi + (1 + \mu|\Psi|^2)\partial_x^2\Psi = 0,
\end{equation}
where $\Psi$ is the continuum correspondence of the complex lattice field amplitude. 
We see that by letting $\mu \rightarrow 0$ we obtain
the nonlinear Schr\"odinger equation, where the sign of $\nu$ determines the character of the
soliton solutions \cite{kivshar-agrawal}, while $\nu \rightarrow 0$ gives a continuum
version of the Ablowitz--Ladik equation \cite{ablowitz-ladik}.

Next, we derive a wave-kinetic equation for the wave function $\Psi$. 
We first introduce the Wigner function for the wave function $\Psi$ according to
\cite{wigner,moyal,mendonca}
%2
\begin{equation}\label{eq:wigner}
  \rho(t,x,k) = \frac{1}{2\pi}\int_{-\infty}^{\infty} d\xi\,e^{ik\xi} 
  \langle\Psi^*(t,x + \xi/2)\Psi(t,x - \xi/2)\rangle ,
\end{equation}
where the angular bracket denotes the ensemble average \cite{klimontovich}. The 
Wigner function is a generalized distribution function for the quasi-particles
representing $\Psi$, and it satisfies the normalization 
%3
\begin{equation}\label{eq:intensity}
  I(t,x) \equiv \langle|\Psi(t,x)|^2\rangle = \int_{-\infty}^{\infty} dk\,\rho(t,x,k) .
\end{equation} 

Applying the time derivative to Eq.\ (\ref{eq:wigner}) and using Eq.\ (\ref{eq:salerno})
we obtain the wave-kinetic equation
%4
\begin{eqnarray}
&&\!\!\!\!\!\!\!\!
  \partial_t\rho + 2k\partial_x\rho 
  + 2I\Big\{ \sin\left( \tfrac{1}{2}\stackrel{\leftarrow}{\partial}_x
    \stackrel{\rightarrow}{\partial}_k \right)\left[ 
    2(\nu + \mu) + \mu(k^2 + \partial_x^2) \right] 
\nonumber \\ &&\qquad
  + \cos\left(  \tfrac{1}{2}\stackrel{\leftarrow}{\partial}_x
    \stackrel{\rightarrow}{\partial}_k \right)\mu k\partial_x \Big\} \rho = 0 ,
\label{eq:kinetic}
\end{eqnarray}
where the $\sin$ and $\cos$ operators are defined in terms of their
respective Taylor expansion, and the arrows denote the direction of operation.
Equation (\ref{eq:kinetic}) describes the propagation of partially
coherent wave modes taking onsite as well as intersite nonlinearities
into account.

Next, we look for unstable perturbations around the homogeneous solution $\rho_0(k)$. 
We let $\rho(t,x,k) = \rho_0(k) + \rho_1(k)\exp(iKx - i\Omega t)$,
where $|\rho_1| \ll \rho_0$, and linearize (4) with respect to $\rho_1$. 
Equation (\ref{eq:kinetic}) with the intensity (\ref{eq:intensity}) then
gives the nonlinear dispersion relation
%5
\begin{equation}\label{eq:disprel}
  1 = \int dk\frac{[2(\nu + \mu) + \mu k^2][\rho_0(k + K/2) - \rho_0(k - K/2)]}
    {\Omega - 2K(1 + \mu I_0)k} .
\end{equation} 
The dispersion relation (\ref{eq:disprel}) generalizes the results of
previous modulational instability analysis of the nonlinear Schr\"odinger 
equation, where the Wigner formalism \cite{anderson-etal} as well as  
the mutual coherence method \cite{demetrios-etal} has been used.

For a coherent background spectrum, i.e.\ $\rho_0(k) = I_0\delta(k)$, the dispersion 
relation (\ref{eq:disprel}) reduces to 
%6
\begin{equation}\label{eq:dispmono}
  \Omega = \pm \Big\{
    (1 + \mu I_0)\left[\left(1 + \tfrac{1}{2}\mu I_0\right)K^4 - 4(\nu + \mu)I_0 K^2\right]
  \Big\}^{1/2} .
\end{equation}
We see that for the nonlinear Schr\"odinger case ($\mu = 0$), we obtain
the standard growth rate $\Gamma = -i\Omega = K[4\nu I_0 - K^2]^{1/2}$,
showing the existence of a modulational instability for focusing ($\nu >0$)
nonlinearity.

For a random phase background wave function, we may represent
the background quasi-particle distribution function by the Lorentz
distribution
%7
\begin{equation}
  \rho_0(k) = \frac{I_0}{\pi}\frac{\Delta}{k^2 + \Delta^2} ,
\end{equation}
where $\Delta$ is the width of the distribution function. Using the Lorentz 
distribution in the nonlinear dispersion relation, we obtain from (6)
%8
\begin{eqnarray}
  &&
  \Omega = -2i\Delta K\left(1 + \tfrac{1}{2}\mu I_0\right) 
  \nonumber \\ &&\qquad
  \pm 
  \Big\{
    (1 + \mu I_0)\left[\left(1 + \tfrac{1}{2}\mu I_0\right)K^4 - 4(\nu + \mu)I_0 K^2\right]
     \nonumber \\ &&\qquad\quad
    + 2\Delta^2K^2\mu I_0\left(1 + \tfrac{1}{2}\mu I_0\right) \Big\}^{1/2} .
\label{eq:displorentz}
\end{eqnarray}
We note that the dispersion relation (\ref{eq:displorentz}) reduces to
(\ref{eq:dispmono}) when $\Delta \rightarrow 0$. Moreover, the case of
a pure Kerr nonlinearity ($\mu = 0$) gives rise to the a linear damping term
for the modulational growth rate. In general, the broadening
of the background quasi-particle distribution is seen to give rise
to a damping of the modulational instability growth rate through the
first term in the dispersion relation (\ref{eq:displorentz}), but through 
the intersite nonlinearity parameter $\mu$ we also have a new coupling
due to the finite width of the distribution function. 

Next, we numerically investigate the properties of the growth rate 
$\Gamma = \mathrm{Im}(\Omega)$, deduced from (8).  In Fig.\ 1 we have plotted 
the modulational instability growth rate for different parameter values. We have 
normalized the parameter $\nu$ to one, and used $\Delta = 0$ and $\Delta = 0.1$. 
For $\mu = \Delta = 0$ we retrieve the standard result for the modulational instability 
growth rate in the case of the nonlinear Schr\"odinger equation. This behavior is, however, 
affected by a finite width in the background wave spectrum.  All cases show a Landau-like 
damping of the growth rate, when a finite width of the background distribution $\rho_0$ is 
introduced.  Thus, appropriate random phasing of the wave function $\Psi$
could act as a useful means of stabilizing, e.g. Bose--Einstein condensates in deep 
optical traps.  It is interesting to note that there is an interplay between the onsite
and intersite nonlinearities through the coefficients $\nu$ and $\mu$, respectively.
As expected, the case of competing nonlinearities, i.e.\ $\mu\nu < 0$ (here
represented by a repulsive intersite nonlinearity),
gives rise to a reduced growth rate as compared to the case of 
a Kerr type nonlinearity ($\mu = 0$), since any perturbation growth
in one nonlinearity can be decreased by the other nonlinear term.  
However, for $\mu, \nu > 0$ (the dotted curves) we note a significant increase
in the growth rate.

%%%%%% FIG %%%%%%
\begin{figure}
\includegraphics[width=0.9\columnwidth]{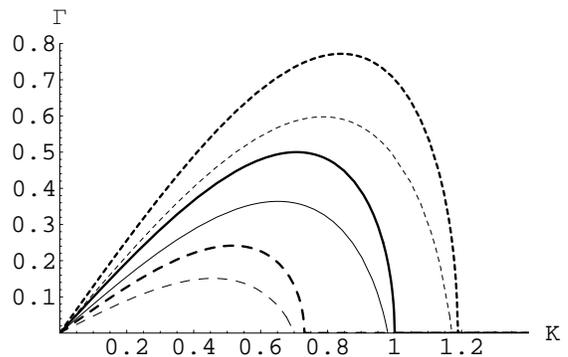}
\caption{The growth rate $\Gamma$ plotted as a function of 
  the perturbation wavenumber $K$ for different parameter values.
  All curves have normalized the parameter $\nu$ to one and intensity
  $I_0 = 0.25$. 
  The heavy solid curve has $\mu = \Delta = 0$, and 
  corresponds to the coherent nonlinear Schr\"odinger modulational instability; 
  the thin solid curve has $\mu =0$ and $\Delta = 0.1$;
  the heavy dotted curve has $\mu = 0.5$ and $\Delta = 0$, while
  the thin dotted curve has $\mu = 0.5$ and $\Delta = 0.1$; finally,
  the heavy dashed curve has $\mu = -0.5$ and $\Delta = 0$, while
  the thin dashed curve has $\mu = -0.5$ and $\Delta = 0.1$. 
  Thus, all cases show damping through the broadening of the 
  background wavenumber distribution.}
\end{figure}
%%%%%%%%%%%%%%%

To summarize, we have performed a wave-kinetic study of the 
Salerno model in the continuum limit. A Vlasov-like equation
has been obtained for the quasi-particles, representing the 
dynamics of the wave function $\Psi$ in phase space. 
A modulational instability analysis has been carried out, and 
a comparison between the coherent and incoherent cases has been made. 
It was found that the interaction between the onsite and
intersite nonlinearities and the finite width of the 
background distribution function gives rise to significant changes
in the modulational instability growth rate.

\acknowledgments
This research was partially supported by the Swedish Research Council as
well as by Centre for Fundamental Physics at the Rutherford Appleton
Laboratory, Chilton, Didcot, United Kingdom.


\begin{thebibliography}{99}

  \bibitem{kivshar-agrawal}
  Yu. Kivshar and G. P. Agrawal, \textit{Optical Solitons, From Fibers to Photonic Crystals} 
  (Academic, 2003). 
  
  \bibitem{shukla-rasmussen}
  P. K. Shukla and J. J. Rasmussen, Opt. Lett. \textbf{11}, 171 (1986).
  
  \bibitem{onorato-etal}
  M. Onorato, A. R. Osborne, and M. Serio, Phys. Rev. Lett. \textbf{96}, 
  014503 (2006). 
  
  \bibitem{hasegawa}
  A. Hasegawa, \textit{Plasma Instabilities and Nonlinear Effects} 
  (Springer-Verlag, New York, 1975).  

  \bibitem{hennig-tsironis}
  D. Henning and G. P. Tsironis, Phys. Rep. \textbf{307}, 333 (1999). 

  \bibitem{salerno}  
  M. Salerno, Phys. Rev. A \textbf{46}, 6856 (1992).
  
  \bibitem{li-etal}
  Z. D. Li, P. B. He, L. Li, J. Q. Liang, and W. M. Liu, Phys. Rev. A \textbf{71}, 053611 (2005)
  
  \bibitem{gomez-gardenes-etal}
  J. Gomez-Garde\~nes, B. A. Malomed, L. M. Flor\'ia, and A. R. Bishop,
  Phys. Rev. E \textbf{73}, 036608 (2006).
    
  \bibitem{ablowitz-ladik}
  M. J. Ablowitz and J. F. Ladik, J. Math. Phys. \textbf{17}, 1011 (1976).
  
  \bibitem{wigner}
  E. P. Wigner, Phys. Rev. \textbf{40}, 749 (1932); 
  
  \bibitem{moyal}
  J. E. Moyal, Proc. Cambridge Philos. Soc. \textbf{45}, 99 (1949). 
  
  \bibitem{mendonca}
  J. T. Mendon\c{c}a, \textit{Theory of Photon Acceleration} (IOP Publishing, Bristol, 2001).
  
  \bibitem{klimontovich}
  Yu. L. Klimontovich, \textit{The Statistical Theory of Non-Equilibrium Processes in a Plasma} 
  (Pergamon Press, Oxford, 1967). 
  
  \bibitem{anderson-etal}
  D. Anderson, B. Hall, M. Lisak and M. Marklund, Phys. Rev. E {\bf 65}, 046417 (2002);
  B. Hall, M. Lisak, D. Anderson, R. Fedele, and V. E. Semenov, {\it ibid.} {\bf 65}, 035602 (2002);
  M. Marklund and P. K. Shukla, Rev. Mod. Phys. {\bf 78}, No. 2, in press (2006).
  
  \bibitem{demetrios-etal}
  M. Soljacic, M. Segev, T. Coskun, D. N. Christodoulides, and A. Vishwanath,
  Phys. Rev. Lett. \textbf{84}, 467 (2000).
  
\end{thebibliography}
\end{document}